\documentclass[%
 aip,
 jmp,%
 amsmath,amssymb,
 reprint,%
]{revtex4-2}

\usepackage{graphicx}% Include figure files
\usepackage{dcolumn}% Align table columns on decimal point
\usepackage{bm}% bold math
\usepackage{amsmath}
\begin{document}

\preprint{AIP/123-QED}

\title[Generation of 18 picosecond keV Ne$\mathrm{^+}$ ion pulses from a cooled supersonic gas beam]{On the path to ion-based pump-probe experiments: \\
Generation of 18 picosecond keV Ne$\mathrm{^+}$ ion pulses from a cooled supersonic gas beam}

\author{L.~Kalkhoff}
 \email{lukas.kalkhoff@uni-due.de}
\author{A.~Golombek}%
\author{M.~Schleberger}%
\author{K.~Sokolowski-Tinten}%
\author{A.~Wucher}%
\author{L.~Breuer}%
\email{lars.breuer@uni-due.de}
 \affiliation{University of Duisburg Essen, Physics Department and CENIDE, Lotharstr. 1, 47057 Duisburg, Germany}

\date{\today}% It is always \today, today,

\begin{abstract}
The dynamics triggered by the impact of an ion onto a solid surface has been explored mainly by theoretical modeling or computer simulation to date. Results indicate that the microscopic non-equilibrium relaxation processes triggered by the interaction of the ion with the solid occur on (sub-)picosecond time scales. A suitable experimental approach to these dynamics therefore requires a pump-probe method with an appropriate time resolution. Recent experiments have successfully used laser photoionization of noble gas atoms in combination with a Wiley-MacLaren ion buncher to obtain arrival time distributions as narrow as $t_{ion}=180$~ps. Here we show that this setup can be significantly improved by replacing the gas at a temperature of $T_{atoms}\simeq 300$~K with a supersonic beam of cooled noble gas atoms at $T_{atoms}\simeq 4$~K. The detailed analysis of measured arrival times of individual Ne$\mathrm{^+}$ ions with a kinetic energy of 4~keV reveals that the arrival time jitter can be reduced by this technique down to (18 $\pm4$)~ps. This opens the door to pump-probe experiments with keV ions with a time-resolution in the picosecond range.

%Which probably manifests the shortest ion pulse ever recorded but measurements suggest it is still limited by the time resolution of the used detection system which is also implied by theoretical trajectory simulation \cite{Kucharczyk.2020, Kucharczyk.2021}.

\end{abstract}

%Use showkeys class option if keyword
\maketitle

\section{\label {Introduction} Introduction:\protect}
Ion irradiation plays a significant role in solid-state physics, industrial applications, and medical treatment. Examples include doping of semiconductors by ion implantation \cite{Ronning.2001, Schmidt.2013, Roccaforte.2022}, fabrication of nanoscale surfaces \cite{Hellborg.2009,Facsko.1999,Rahali.2023}, shaping of nanopores in two-dimensional systems such as $\mathrm{MoS_2}$ or graphene oxide \cite{Foller.2022,Kozubek.2019,Grossek.2022}, analysis of interlayers in lithium-ion batteries \cite{Zhou.2020}, or the radiotherapy of cancer \cite{Miyamoto.2007, Suit.2010, Tsujii.2012}. 

Despite the widespread use of ion beams, the understanding of the underlying interaction mechanisms is still incomplete. To this date, studies on the interaction of ions with solids are basically limited to the observation of the final state of the dynamic processes triggered by the ion impact. A few attempts have yet been made to use in-situ spectroscopy to investigate the processes. Herder \cite{Herder.2020} and Skopinski \cite{Skopinski.2023}, e.g., used secondary ion/neutral mass spectrometry to determine the velocity distribution of sputtered particles. Both found indications for thermally driven mechanisms but also these experiments target processes taking place rather at the end of the interaction and their interpretation remains difficult. Because the ion-solid dynamics has not been experimentally accessible, it has been studied exclusively by computer simulations and theoretical approaches. These models come to the conclusion that the time frame, where most of the interactions take place, is less than a few picoseconds \cite{Harrison.1988,Lindenblatt.2006, A.Wucher.2013}. The experimental challenge is thus to time-resolve the dynamics triggered by the ion impact, which results e.g. in particle emission and ion formation.

So far, the processes within the solid after an ion impact have not been directly observable, since an experimental approach would require a pump-probe\cite{Fischer.2016} experiment with a time resolution of a few ten picoseconds or less\cite{Gruber.2016}. The realisation of such an experiment is highly attractive and several groups are currently moving into this direction (see e.g. \cite{Golibrzuch.2022}). However, while nowadays light pulses with pulse widths down to the attosecond range are routinely generated \cite{Chini.2014, Gaumnitz.2017}, for keV ions pulses with duration of just below 1~ns could be generated \cite{Hohr.2008}. This \textit{magic barrier} of $\approx1$~ns is due to the initial kinetic energy distribution of the neutral particles, i.e. their temperature if created from a gas, and broadening by space charge effects after the ion creation.

In our first attempt to enable a higher temporal resolution, we were able to reduce the arrival time jitter of mono-energetic 2~keV Argon ions to below 180~ps \cite{Golombek.2021}. Our approach was based on the idea to use a fs~laser for the generation of ions from a neutral gas volume and bunching them within an ion-optical buncher, as reported by Breuers et al\cite{Breuers.2019}.

To establish a pump-probe experiment with even better temporal resolution in the few ps range it was demonstrated by simulations \cite{Kucharczyk.2020, Kucharczyk.2021} that the neutral gas had to be cooled before the ionization. In this paper we experimentally validate this approach by demonstrating arrival time distributions of single keV ions with a width in the 10 ps range.. To this end our previous experiment has been altered to include a geometrically cooled supersonic beam crossed with a fs-laser pulse within an improved adjustable buncher setup. Furthermore, because Argon tends to form clusters of several atoms at low temperatures, we switched to Neon which is more suitable for this application. As we will discuss below, this supersonic beam has a perpendicular temperature of $T_{\perp}=3.6$~K which leads to a record low value for the arrival time jitter of mono-energetic Ne ion pulses of $\left(18\pm 4\right)$~ps full width at half maximum (FWHM).
Please note, that for simplicity we will refer from hereon to the width of the arrival time distribution as an (effective) ion pulse duration. This is motivated by results of our recent theoretical analysis\cite{Kucharczyk.2020}. By including both, the arrival time jitter as well space charge effects, we were able to show, that with our cooling approach it should be possible to generate short pulses containing up to 100 ions. 
 
\section{\label{Methods} Experimental setup \protect}

\begin{figure}[ht]
       \centering
       \includegraphics[width=0.8\textwidth]{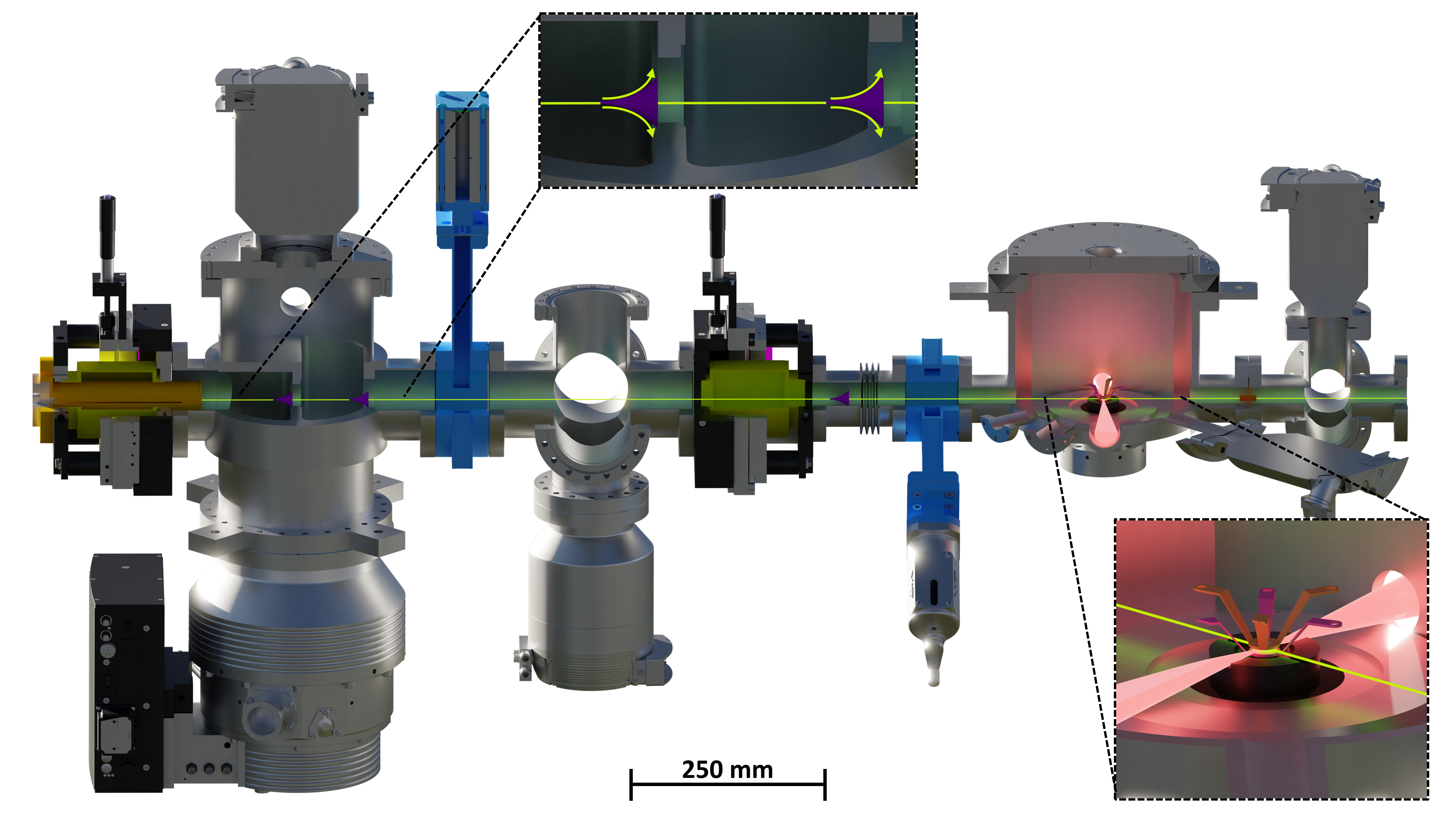}
   \caption{3D rendering of the chamber system for the generation of a geometrically cooled supersonic beam and picosecond ion pulses. An ultrasonic nozzle generates a supersonic beam (green) in the expansion chamber on the left which is led through three skimmers as seen in the inset on the top. Gas atoms with a non-zero velocity component perpendicular to the beam axis (green arrows) will be cut off before ionization when the supersonic beam is crossed with the red marked laser beam within the ion buncher (see also Fig.\ref{Buncher Schema}) as seen on the bottom right inset.}
    \label{Gasbeam Schema}
\end{figure}  

The basic concept for ion pulse generation includes the photoionization of atoms by a laser beam, which are then accelerated towards a target. The main limiting factor for the achievable temporal resolution (effective ion pulse duration) in a pump-probe type experiment is the variation of the arrival time of individual ions at the target due to their initial velocity distribution in the ionization volume. Different starting velocities in the direction of the extraction axis of the ion, e.g. due to thermal broadening, lead to different flight times and thus a broad arrival time distribution, i.e. long effective pulse duration. To accomplish ion pulse widths below 50 picoseconds, the gas must be cooled \cite{Kucharczyk.2020}. Therefore, we equipped our ultrahigh vacuum (UHV) chamber at $\mathrm{3 \times 10^{-8}}$~mbar with a 1.4~m long section, where a geometrically cooled beam of neutral atoms is produced, as shown in Fig.~\ref{Gasbeam Schema}. 

Starting from a supersonic gas valve (Amsterdam Piezo Valve, MassSpecpecD BV), a supersonic beam is created in the first UHV chamber, so called expansion chamber, from a reservoir of Neon (Ne) at 2 bar. This beam can be formed in a DC like manner or in a pulsed mode to adjust the gas density. Because of the supersonic gas expansion under UHV conditions\cite{Luria.2011}, at a certain distance from the nozzle, the atoms cannot interact with each other anymore, which will lead to frozen trajectories of said atoms. This distance is called the freezing plane described by the \textit{Sudden-Freeze}-model\cite{DePonte.2006, Habets.1977,Beijerinck.1981}. Gas atoms with a large velocity component perpendicular to the supersonic beam axis can therefore be cut off behind the freezing plane with so called skimmers, as shown in Fig.~\ref{Gasbeam Schema}, upper inset. The supersonic beam is directed through these skimmer's pinhole and gas atoms with a non-zero velocity perpendicular to supersonic beam are peeled off as the green arrows indicate. In this setup we use two skimmers with a 1.5~mm pinhole in the expansion chamber and one skimmer with a 1~mm pinhole before the ionisation chamber.

To achieve UHV conditions throughout the system, the expansion chamber is pumped by a Leybold MAG Integra turbomolecular pump (TMP) with pumping speeds of 1000~$\mathrm{\frac{l}{s}}$, the second and the third chamber is equipped with two Leybold TurboVac 350 iX TMPs with pumping speeds of 360~$\mathrm{\frac{l}{s}}$. The supersonic beam is then fed into a beam dump chamber with a narrow opening to reduce the impact of back scattering of non-ionized atoms.

In the ionisation chamber, the supersonic beam is then crossed by a perpendicular laser beam focused down to a spot size of 10~$\mathrm{\mu m}$ (FWHM) by an off-axis parabolic mirror ($f$~=~101.6~mm) in the center of the supersonic gas beam. The laser light used is of 800 nanometers wavelength from a tabletop laser system (Coherent Legend Elite) which delivers 50~fs laser pulses at 1~kHz repetition rate. 
\begin{figure}[h]
    \centering
   \includegraphics[width=0.65\textwidth]{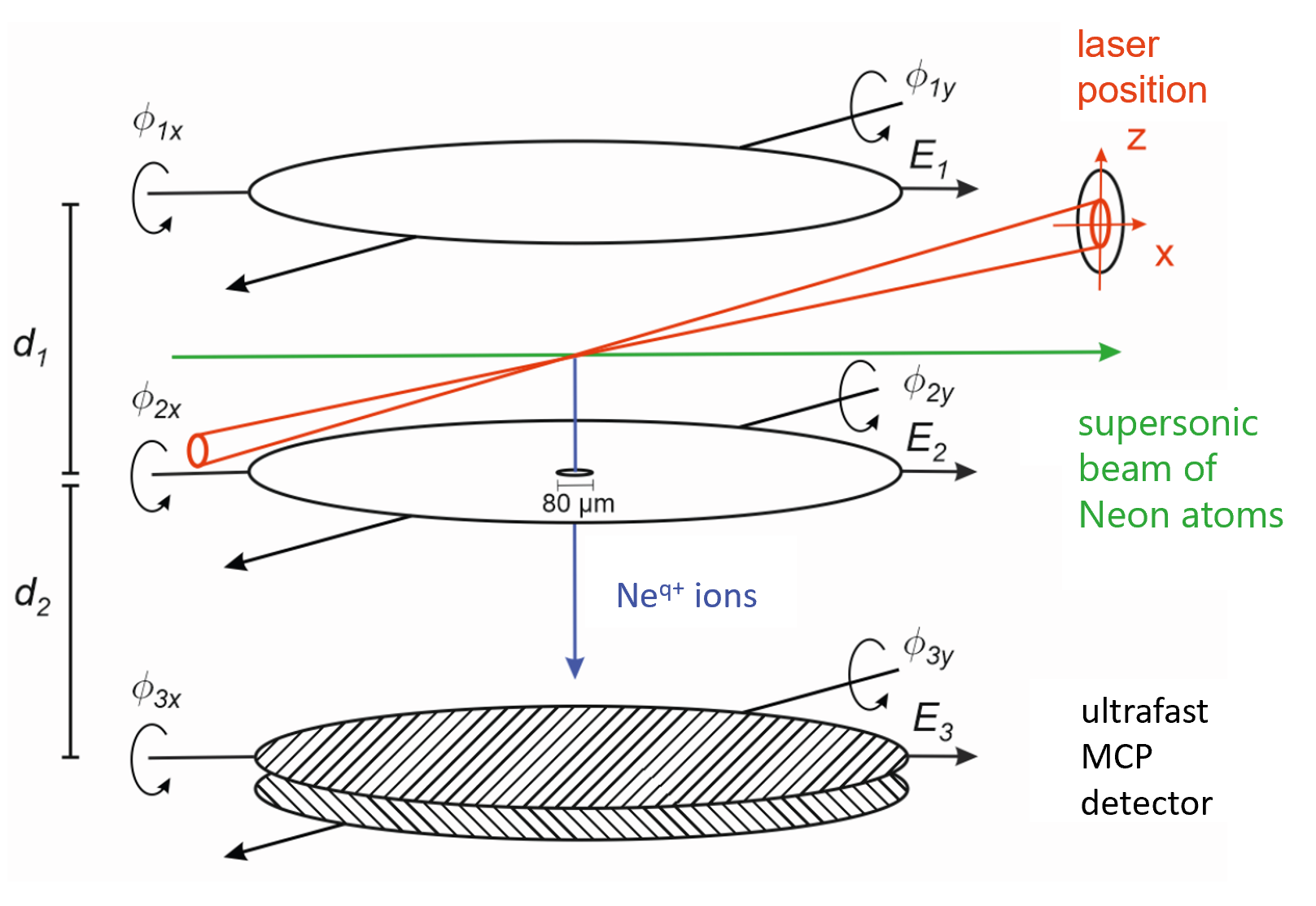}
    \caption{Schematic of the ion buncher. The laser (red) is focused in the center of the Ne supersonic beam (green) between two electrodes $E_1$ and $E_2$ which are at a distance of $d_1$. An electric field between those plates accelerates ions towards the pin hole in $E_2$, which therefore travel the distance $d_2$ towards the ultrafast MCP detector. Every segment of the buncher can be moved independently in height and rotated by $\phi_x$ and $\phi_y$ with respect to the Ne supersonic beam for an optimal alignment\cite{Golombek.2022}}
   \label{Buncher Schema}
\end{figure}
The neutral atoms from the supersonic beam are ionized via tunnel ionisation due to the high electromagnetic field strength in the focal point of the laser ($\approx10^{16}$~W/cm$^2$). Under these conditions, the Coulomb potential of a gas atom is bent to such a degree that electrons can tunnel through the potential barrier. The Keldysh parameter $\gamma$ indicates the possibility of such tunnel ionization process. For $\gamma<<1$ tunnel ionization is more likely to occur and if $\gamma>>1$ multiphoton absorption is the cause for the ionization \cite{Keldysh.1964}. For our setup with Ne$\mathrm{^+}$ ions this parameter is $\gamma = \sqrt{IP/2U_p}=0.53$ with an ionization potential ($IP$) of 14.53~eV and $U_p$ as ponderomotive potential of an electron in the used laser field. Therefore, with a $\gamma<1$, tunnel ionization is the dominant process in our case \cite{Augst.1989}. A $\mathrm{\lambda/2}$ plate in combination with a thin film polarizer allows adjusting the laser intensity before entering the chamber.
 
The intersection of the laser and supersonic beam lies between an electrode arrangement that is called \textit{ion buncher}, see Fig \ref{Buncher Schema}. The electrodes $E_1$ and $E_2$ can be put on a potential difference of (0 -10)~kV with respect to ground and the ions will be accelerated towards the electrode $E_2$. A small hole at the center of $E_2$ with a diameter of 80~$\mathrm{\mu m}$ enables the extraction of ions. The ions then move through a field free area towards a microchannel plate (MCP) detector (Photonis Gen2 ToF Detector). This MCP is specifically designed for a high time resolution by its small incidence angle of 12$\mathrm{^\circ}$ and reduced opening of a single channel of 2~$\mathrm{\mu m}$. Such a three electrode setup is also known as a Wiley-McLaren configuration\cite{Wiley.1955}, where in our case the third electrode is the front of the MCP detector. Due to the field-free area between $E_2$ and the MCP front, the total time-of-flight of ions generated at $z_0=\frac{d_2}{2}$ is given by:

\begin{equation}
\centering
t=\sqrt{\frac{md_1}{qe\Delta \varphi}} \cdot \biggl(\sqrt{2z_0}+\frac{d_2}{\sqrt{2z_0}}\biggr)
\label{eq1}
\end{equation}

where $m$ is the mass of the ion, $q$ the charge state, and $e$ the elementary charge. $\Delta \varphi$ is the total potential difference between both electrodes $E_1$ and $E_2$. Different starting positions generated within the laser focus will now be \textit{bunched} to a time-focus where all ions will hit the MCP front (or the target, respectively) at the exact same time and result in a minimal effective pulse width. To achieve this time-focusing at the MCP, it is mounted on an adjustable flexible flange and the laser focus position can be moved independently. For maximum planarity between each segment, the electrodes and the MCP can be moved separately via piezo motors (SmarAct SLC-1720) with respect to height and angle with a precision of less than $\leq$ 1~$\mathrm{\mu m}$ and $\leq$ 0.1$^{\circ}$, respectively. 

A single ion hitting the MCP will generate an output pulse of 1-20~mV at the 50~$\mathrm{Ohm}$ termination of the collector. The signal is then fed into a constant fraction discriminator (CFD, \textit{Surface Concept Stand-Alone Preamplifier CFD Combination}) for an improved time resolution due to the pulse height statistic given by the nature of an MCP. The stabilized output signal of the CFD is then transferred to a time-to-digital converter (TDC, \textit{Surface Concept TDC-1000}), with a minimal digital bin size of 27.4~ps. The TDC compares the signal from the CFD with the starting signal generated by a small fraction of the laser hitting a fast avalanche photo diode (BPW-28) before entering the ionization chamber.

 Due to the limited input pulse separation of a few nanoseconds given by the TDC, only a single ion of one species can be detected per laser pulse. The output signal of the TDC is then displayed in a histogram to determine the arrival time of each ion species onto the MCP and from the FWHM we determine the resulting effective ion pulse widths. 

\section{\label{Results} Results \protect}

\begin{figure}[h]
    \centering
      \includegraphics[width=0.85\textwidth]{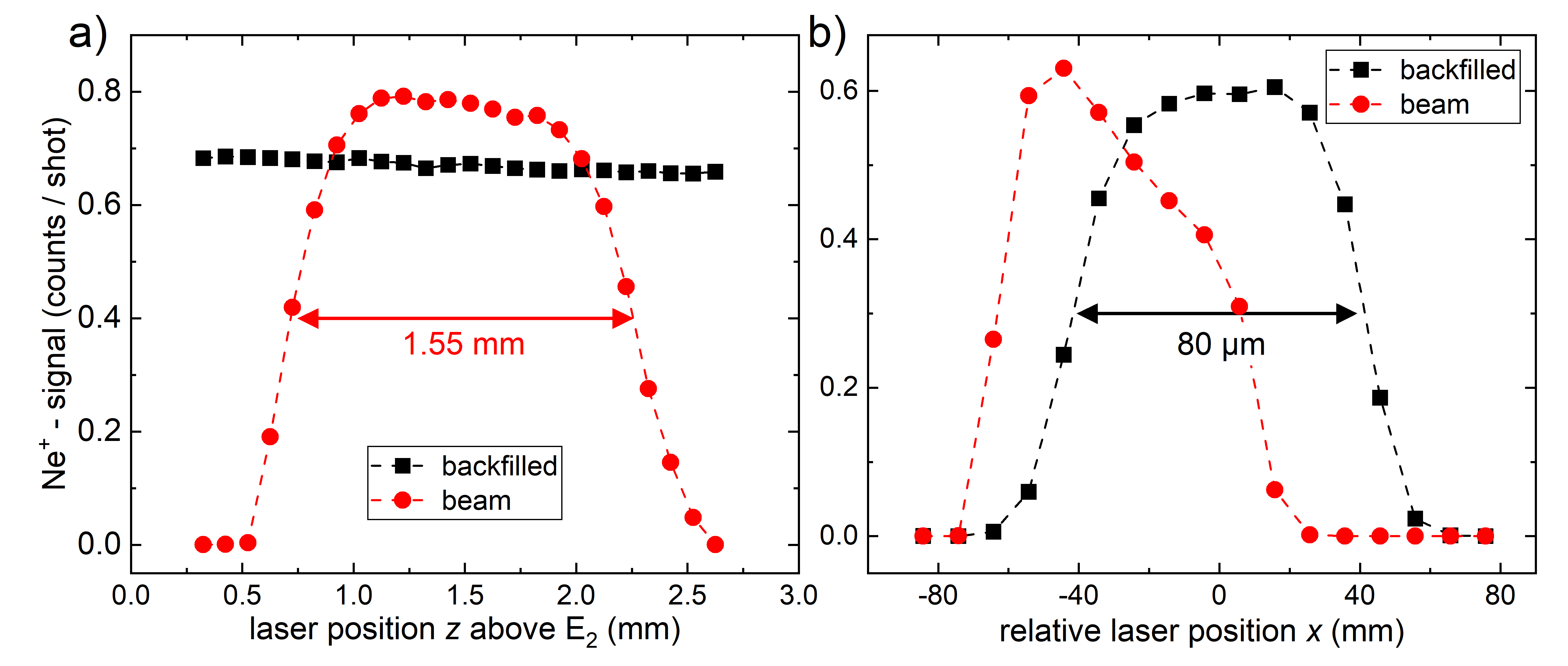}
    \caption{Ne$\mathrm{^+}$ signal of the backfilled Ne (black) to $\mathrm{1.0 \times 10^{-7}}$~mbar and the Ne beam (red) with variation of laser focus position. a) shows a laser scan in $z$-axis between $E_1$ and $E_2$. The supersonic beam can be measured with a diameter of 1.55~mm at the point of ionization, which is not seen in the backfilled gas signal due to homogeneous distribution of the gas atoms. b) shows the variation of the $x$-coordinate and the 80~$\mathrm{\mu m}$ pin hole in $E_2$. A shift of the maximum signal in the beam signal of 20~$\mathrm{\mu m}$ compared to the backfilled signal is given through the parabolic trajectories after the ionization. }
    \label{Geometrie}
\end{figure}

With a successful alignment of supersonic beam and laser beam, Ne ions can be detected by moving the laser focus along the extraction direction, i.e.~the $z$-axis. By this experiment the \textit{z}-dimension, which represents the diameter of the supersonic beam, can be determined. The result is depicted in Fig.~\ref{Geometrie} a) by the red data points. The supersonic beam has a diameter of roughly 1.55~mm in the center of the ion buncher. This result is compared with the chamber filled with Ne gas at $\mathrm{1.0 \times 10^{-7}}$~mbar as depicted in Fig. \ref{Geometrie} a) by the black data points, which will from now on be called \textit{backfilled} Ne gas. The latter shows no dependence of the signal on the laser position, due to the homogeneous density distribution within the vacuum chamber. 

Next, the laser focus was moved across the $x$-axis, which is parallel to the electrode surface. Like in previous work\cite{Golombek.2021}, the aperture of 80~$\mathrm{\mu m}$ in the second electrode can be identified as depicted in \ref{Geometrie} b) by the black data points for the backfilled Ne gas. This shape is slightly deformed and displaced for the Ne supersonic beam, due to the parabolic trajectories of the ions after ionization. This shape is caused by the forward component along the neutral supersonic beam trajectory of the velocity vector. Most of the ions generated above the center of the pinhole in $E_2$ will be deflected onto the electrode surface. Therefore, the center signal for the Ne supersonic beam shifts roughly 20~$\mathrm{\mu m}$ along the supersonic beam direction. Due to the fact, that the 80~$\mathrm{\mu m}$ hole is much smaller than the $x$-dimension of the supersonic beam, the whole beam width cannot be mapped.
\begin{figure}[h]
    \centering
      \includegraphics[width=0.8\textwidth]{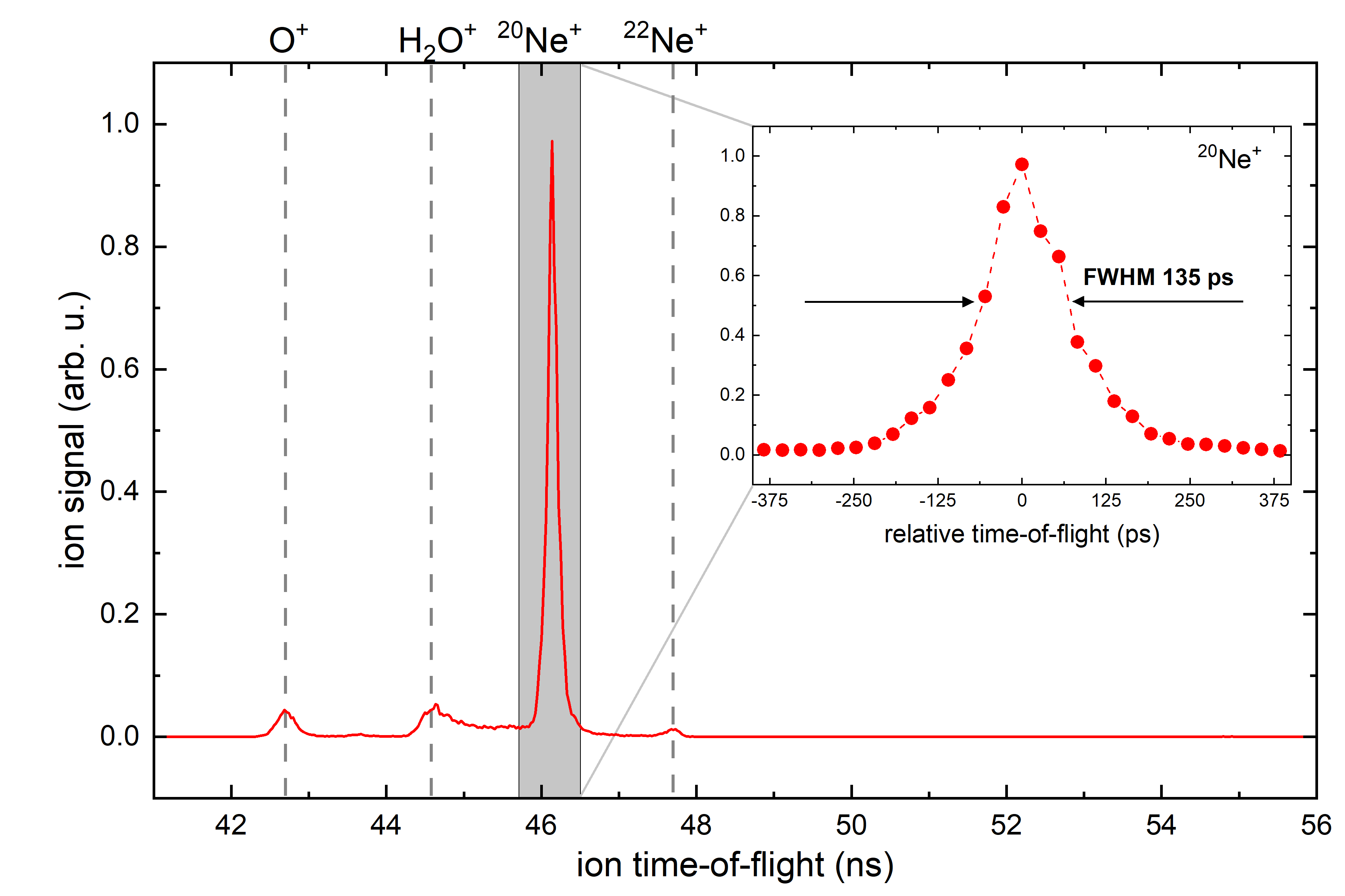}
    \caption{Time-of-flight measurement of ionized gas atoms with the supersonic beam active with 2~keV kinetic energy. Components in the supersonic beam consist of $\mathrm{^{20}Ne^+}$, $\mathrm{^{22}Ne^+}$, whereas $\mathrm{O^+}$, $\mathrm{H_2O^+}$ residual gas is seen even without the supersonic beam active. The inset shows the $\mathrm{^{20}Ne^+}$ signal magnified and the effective pulse width FWHM of 135~ps.}
    \label{Spektrum}
\end{figure}
In Fig.~\ref{Spektrum} an example is shown of the a time-of-flight spectrum with the Ne supersonic beam active. With the given laser power of $\mathrm{1x10^{16}~\frac{W}{cm^2}}$, singly charged oxygen, water and Ne ions can be seen. Also, the isotope $^{22}$Ne can be seen in the given spectrum. All ions are generated with 2~keV kinetic energy.

\begin{figure}[h]
    \centering
      \includegraphics[width=0.9\textwidth]{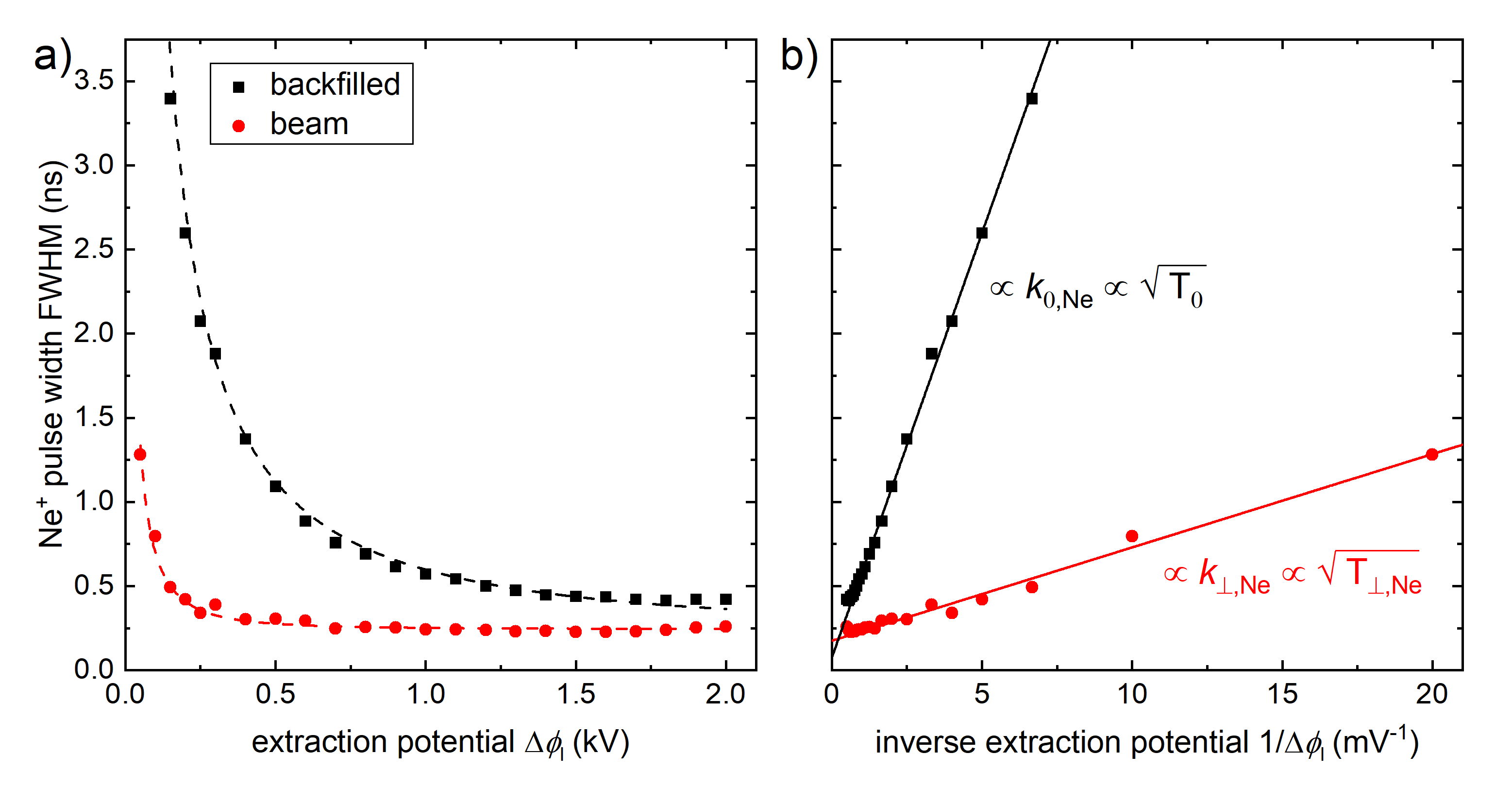}
    \caption{Pulse width dependency of the $\mathrm{Ne^+}$ ions over the extraction potential between the two buncher electrodes. Black dots represent data taken with the chamber backfilled with Ne gas and the red dots represent data taken with ions produced from the supersonic beam. In a) the dashed line guides the eye and shows at lower extraction potentials a strong dependency of the pulse width, which reaches the resolution limit at different potentials depending on the source of the ion. In b) the extraction potential is inversed to show the linear dependency of pulse width with $\mathrm{\sqrt{T}}$, confirmed by the linear fit.}
    \label{Temperatur}
\end{figure}

The FWHM of the Ne signal can be seen in the inset of Fig.~\ref{Spektrum} with an effective pulse width of 135~ps. Even though it is a shorter pulse width in comparison to previous work performed by Golombek et. al with 180~ps\cite{Golombek.2021} for backfilled Argon ions, the result is much higher than theory predicts for ions with lower temperature. To further investigate the reason, the extraction potential is increased, which results in the same pulse width of 135~ps. This behaviour is contradictory to previous work and theory, as the pulse width should decrease with increasing extraction potential, indicating a fundamental problem. Fig.~\ref{Temperatur} a) shows the pulse width of $\mathrm{Ne^+}$ ions over the extraction potential from 0 to 2~keV for the supersonic beam represented by the red data points and for backfilled Ne gas in black. For the latter, the expected behaviour of the pulse width is clearly seen, the results from the supersonic beam show no further dependency from 0.5~keV to 2.0~keV. This leads us to the hypothesis that we reached the detection limit of our set-up and can thus not directly detect pulse widths below 135~ps.   

To get a grasp of the true ion pulse width, even with the limited time resolution of the MCP detection system, we varied the extraction potential to further investigate the dependency of the effective pulse width and to estimate the temperature of the gas atoms in the supersonic beam. By decreasing the potential between $E_1$ and $E_2$, the thermal broadening effect is less and less compensated by the accelerating field, resulting in broader effective pulse widths with decreasing potential. The dependency of the pulse width on the temperature of the atoms and the accelerating field, respectively, are described in detail in Refs. \cite{Kucharczyk.2020, Kucharczyk.2021,Golombek.2022}. In short, we need to calculate the velocity distribution perpendicular to the beam described by a Boltzmann distribution and its standard deviation $\sigma_{v_z}$: $\Delta v_z = 2\sqrt{2ln(2)} \cdot \sigma_{v_z}$, the accelerating force given by $ \frac{\partial v_z}{\partial t(v_z)}= q e \Delta \varphi$ where $q$ is the charge state of the ion, $e$ the elementary charge and $\Delta \varphi$ expresses the potential difference of $E_1$ and $E_2$. Therefore:   

\begin{equation} \label{eq2}
   \Delta t_{v_z}  = \frac{\partial t(v_z)}{\partial v_z} \cdot \Delta v_z = 2\sqrt{2ln(2)} \cdot \sqrt{\frac{k_B T}{m}} \cdot \frac{m d_1}{q e \Delta \varphi} 
\end{equation}

Where $m$ is the mass of the atom, $k_B$ the Boltzmann constant, $T$ is the temperature in Kelvin and $d_1$ the distance of the first two buncher electrodes. Therefore, the pulse width is proportional to the square root of $T$: 
\begin{equation} \label{eq3}
    \Delta t_{v_z} \propto \frac{\sqrt{T}}{\Delta \varphi}
\end{equation}

 With higher potential differences between the buncher electrodes, a decrease of the pulse width should be observed as can be seen from equation \ref{eq3}. And this is indeed observed in the experimental data from the backfilled Ne gas, but in case of the supersonic beam we do not see such a behaviour for the full range of the extraction potential. The point where the pulse width of Ne from the supersonic beam levels off, is at $\Delta \varphi_1=0.3$~kV. This is the time resolution limit given by the electronics used in this experiment. With equation \ref{eq3}, we can thus estimate a temperature for the supersonic beam. The difference between backfilled and supersonic gas beam is emphasized in figure \ref{Temperatur} b), where the data is plotted over the inverse of $\Delta \varphi_1$ and can be fitted by a linear function where the slope is proportional to $\sqrt{T}$. Assuming that the backfilled Ne is at $T_0$~=~300~K we get: 

\begin{equation} \label{eq4} 
T_{\perp,Ne} = \biggl(\frac{k_{\perp,Ne}}{k_{0,Ne}}\biggr)^2 \cdot T_0  = 3.6~\mathrm{K}
\end{equation}

Thus, we have shown that the geometrical cooling is effective and that the cooled supersonic beam has a perpendicular temperature of $T_{\perp}=3.6$~K. With this conclusion in mind we can now estimate our effective pulse width, by following the concept of equation \ref{eq2}. With the temperatures $T_{\perp,Ne}=3.6$~K, $T_{\perp,0}=300$~K and $t_0=175$~ps, we get: 

\begin{equation}
    \Delta t\leq\sqrt{\frac{T_{\perp,Ne}}{T_0}} \cdot \Delta t_0 = (18 \pm 4)~\mathrm{ps}
\end{equation}

This is the shortest ion pulse ever recorded. We would like to point out, that this value is still an upper limit and that the true pulse width should be ideally determined by a direct measurement which would however require a faster detection scheme, such as a streak camera, for example.

\section{\label{Conclusion} Conclusion and outlook \protect}
The results presented here show, that it is in principle possible to overcome the magic barrier of a few hundred picoseconds for keV ion pulses. Further, we demonstrated that the cooled supersonic beam is a key component for our setup. With T = 3.6 K we obtain an upper boundary of the effective ion pulse width of about 18 ps, the shortest ion pulses in the keV-range we are aware of and at least an order of magnitude shorter than what we achieved before without cooling. With this we have completed the most important step for a pump-probe experiment using ions with energies in the keV range. Since these ions are generated by photoionization, the ion pulses are synchronized with the fs laser. Thus, a second laser pulse, split from the main laser beam and optically delayed to compensate for the travel time of the ion from its point of generation to the surface, can act as a probe. With these two pillars the first pump-probe experiment with synchronized photons and monoenergetic ions in the keV regime can be realised with a time resolution of about 10~picoseconds.

\begin{acknowledgments}
The authors are greatly indebted to Ping Zhou for his assistance with the laser system. We thank the Deutsche Forschungsgemeinschaft (DFG) for their financial support of project C05 within the Collaborative Research Center (CRC) 1242 ‘Non-equilibrium dynamics in the time domain’ (project number 278162697).
\end{acknowledgments}

\nocite{*}
\bibliography{18ps_ion_pulses}% Produces the bibliography via BibTeX.

\end{document}